# Oxygen abundance methods in the SDSS: view from modern statistics

F. Shi, G. Zhao, James Wicker

National Astronomical Observatories, Chinese Academy of Sciences, 20A Datun Road, Chaoyang District, Beijing 100012, PR China
e-mail: fshi@bao.ac.cn

**ABSTRACT**

*Context.*
*Aims.* Our purpose is to find which is the most reliable one among various oxygen abundance determination methods.
*Methods.* We will test the validity of several different oxygen abundance determination methods using methods of modern statistics. These methods include Bayesian analysis and information scoring. We will analyze a sample of ∼6000 H II galaxies from the Sloan Digital Sky Survey (SDSS) spectroscopic observations data release four.
*Results.* All methods that we used drew the same conclusion that the $T_e$ method is a more reliable oxygen abundance determination methods than the Bayesian metallcity method under the existing telescope ability. The ratios of the likelihoods between the different kinds of methods tell us that the $T_e$, $P$, and $O3N2$ methods are consistent with each other because the $P$ and $O3N2$ method are calibrated by $T_e$-method. The Bayesian and $R_{23}$ method are consistent with each other because both are calibrated by a galaxy model. In either case, the $N2$ method is an *unreliable method*.

**Key words.** galaxies: abundance – galaxies: starburst – stars: formation

## 1. Introduction

H II galaxies with strong emission lines are important probes for the formation and evolution of galaxies. Their spectra contain much important information needed to determine the star formation rate, initial mass function, element abundance, etc. (Stasińska & Leitherer 1996; Kennicutt 1998; Contini et al. 2002). The heavy element abundance is a key parameter for the formation and evolution of a galaxy. Oxygen is one of the most important elements and is most easily and reliably determined since the most important ionization stages can all be observed.

There are various methods for determining the oxygen abundance for the H II galaxies based on the strong emission lines. The oxygen abundance by the measurement of electron temperature from [O III]$\lambda\lambda$4959,5007/[O III]$\lambda$4363 is one of the most reliable methods. Tremonti et al. (2004) provided Bayesian abundances statistically by the MPA/JHU group [1], based on simultaneous fits of all the most prominent emission lines ([O II], H$\alpha$, [O III], H$\beta$, [N II], [S II]) with a model designed for the interpretation of integrated galaxy spectra (Charlot & Longhetti 2001). Instead of the $T_e$ method and Bayesian method, strong line methods such as the $R_{23}$ [2], $P$ [3], $N2$ [4] or $O3N2$ [5] methods are used widely (Pagel et al. 1979; Kobulnicky et al. 1999; Pilyugin et al. 2001; Charlot & Longhetti 2001; Denicoló et al. 2002; Pettini & Pagel 2004; Tremonti et al. 2004; Liang et al. 2006). Among these methods, which one is most reliable is an unsolved problem and debated broadly by many investigators (Stasińska 2005; Pilyugin & Thuan 2005; Shi et al. 2005, 2006, 2007; Kobulnicky et al. 1999). So it is necessary to study this issue again and find the most reliable oxygen abundance determination method.

The judgment of which method is the best one depends on how well a model agrees with the data. If one wants to fit the data better, expanding the set of free parameters in the model is needed. Thus, adding more free parameters will improve the fit, but make the model more complex. A more complex model is unsatisfactory compared to a simple model if the two models fit the data equally well. Therefore, one must decide what improvement of accuracy in the fit and the penalty paid by introducing the new parameter.

Such consideration forms the spirit of Occam's razor [6]. A quantitative formulation should combine the goodness of fit with a penalty function measuring the complexity of the theory, or directly measure the credibility of the model rather than the frequency of occurrence as in the classical approach. Bayesian statistical analysis and information scored statistical analysis try to find the most parsimonious models that adequately fit the data. These approaches grow from the requirements mandated by Occam's razor. Under Bayesian statistical methods, this permits us to assign a posterior probability for the validity of a physical model, in our case of the method of calculating the oxygen abundance. A method is preferred when its posterior probability exceeds that of any other competing method.

The Sloan Digital Sky Survey (SDSS) is the most ambitious imaging and spectroscopic survey to date, and will eventually

---

*Send offprint requests to*: F. Shi

[1] See http://www.mpa-garching.mpg.de/SDSS/.
[2] $R_{23}$=([O II]$\lambda$3727+[O III]$\lambda\lambda$4959,5007)/H$\beta$
[3] $P$=[O III]$\lambda\lambda$4959,5007/([O II]$\lambda$3727+[O III]$\lambda\lambda$4959,5007)
[4] $N2$=log([N II]$\lambda$6583/H$\alpha$)
[5] $O3N2$=log(([O III]$\lambda$5007/H$\beta$)/([N II]$\lambda$6583/H$\alpha$))

[6] Occam's razor is a principle attributed to the 14th-century English logician and Franciscan friar William of Ockham. Occam's razor is often paraphrased as "All things being equal, the simplest solution tends to be the best one."

cover a quarter of the sky (York et al. 2000). The large area coverage and moderately deep survey limit of the SDSS make it suitable for studying the physical properties of the galaxy. Because of its good homogeneity, the SDSS provides a large sample of H II galaxies where oxygen abundance can be calculated with the various methods.

This paper is organized as follows: based on a SDSS DR4 starbursts spectral sample, we present a sample which we can use to calculate the oxygen abundance with various methods (Sect.2). In Sect. 3, we calculate the oxygen abundance with various methods. In Sect. 4, we measure the credibility of various oxygen abundance determination methods using Bayesian analysis and information scoring. In Sect. 5, we discuss the reliability of various oxygen abundance determination methods, and conclude this paper.

## 2. Data sample

We have used H II galaxies from the Fourth Data Release (DR4) of the SDSS. After subtracting the underlying starlight using the method of Li et al. (2005) and Lu et al. (2006), we fit the emission lines using the method of Dong et al. (2005). We made the internal reddening correction for the flux of all the emission lines, using the two strongest Balmer lines, H$\alpha$/H$\beta$ and the effective absorption curve $\tau_\lambda = \tau_V(\lambda/5500\text{Å})^{-0.7}$, which was introduced by Charlot & Fall (2000). Then, we made use of the spectral diagnostic diagrams from Kauffmann et al. (2003) to classify galaxies as either starburst galaxies, active galactic nuclei (AGN), or unclassified. To reduce systematic and random errors from aperture effects, our galaxy sample is limited by the requirement of redshift $z > 0.04$ (Kewley et al. 2005).

Within the primary starburst sample, two subsamples were selected from the SDSS-DR4 with the fluxes of spectral lines for all [O II]$\lambda$3727, H$\beta$, [O III]$\lambda$4959, [O III]$\lambda$5007, H$\alpha$ and [N II]$\lambda$6583 higher than the flux uncertainty. The difference between these two subsamples is that the first subsample (**Sample I**) is selected by the additional criterion from the [O III]$\lambda$4363 line, that the flux uncertainties for [O III]$\lambda$4363 are higher than $1\sigma$. 409 galaxies were included in this subsample. [O III]$\lambda$4363 is strongly dependent on the metallicity of galaxies; it becomes undetectable in high metallicity galaxies. Therefore, galaxies in Sample I are low metallicity galaxies. In the second subsample (**Sample II**), galaxies have weak or no [O III]$\lambda$4363 line. 5880 galaxies were selected in this subsample. The average uncertainty in flux measurement in the computed $12 + \log(\text{O/H})$ values is typically 0.10 dex in both of Sample I and Sample II.

## 3. Determination of oxygen abundance

### 3.1. $T_e$ method

To derive oxygen abundances with the $T_e$ method, we determined $T_e$ and $n_e$ for a two-zone photoionized H II region model. It is well established that for low metallicity galaxies, $12 + \log(\text{O/H}) < 8.2$, [O III]$\lambda$4363 is prominent and can be measured accurately, while for high metallicity galaxies, [O III]$\lambda$4363 is weak and the error of its measurement is large. We use an $N2$ indicator to distinguish high metallicity regions from low metallicity regions (see Sect. 3.4). For low metallicity galaxies $(12 + \log(\text{O/H})_{N2} < 8.2)$, we used a five-level statistical equilibrium model in the IRAF NEBULAR package (de Robertis, Dufour, & Hunt 1987; Shaw & Dufour 1995), which made use of the latest collision strengths and radiative transition probabilities to determine the $T_e$ and $n_e$. For high metallicity galaxies $(12 + \log(\text{O/H})_{N2} > 8.2)$, an empirical relation of $T_e$ and strong spectral lines has been adopted for the electron temperature determination (Pilyugin 2001). The temperature will be used to derive the $O^{+2}$ ionic abundances.

To estimate the temperature in the low-temperature zone $T_e(\text{O II})$, the relation between $T_e(\text{O II})$ and $T_e(\text{O III})$ from Garnett (1992) was utilized:

$$t_e(\text{O II}) = 0.7 \times t_e(\text{O III}) + 0.3, \quad (1)$$

where $t_e = T_e/10^4$ K. The temperature $T_e(\text{O II})$ is used to derive the $O^+$ ionic abundance.

After calculation of $T_e$ and $n_e$ for the high-temperature zone and low-temperature zone, we used the expressions from Izotov et al. (2006) to calculate the oxygen abundance for these two zones. Then we simply sum $O^+$ and $O^{++}$ as our final oxygen abundance.

### 3.2. $R_{23}$ method

$T_e$ metallicity determination requires the accurate measurement of the weak auroral forbidden emission line [O III]$\lambda$4363. The flux intensity of [O III]$\lambda$4363 strongly anticorrelates with the abundance of galaxies. Its flux intensity becomes undetectable in high metallicity galaxies $(12 + \log(\text{O/H}) > 8.2)$.

For this reason, the strong line metallicity indicator $R_{23}$ has been developed since Pagel et al. (1979) introduced it for the first time. We use the most recent $R_{23}$ analytical calibrations given by Kobulnicky et al. (1999) which are based on the models by McGaugh (1991) to determine the oxygen abundances in our sample.

The major difficulty associated with this method is that the relation between oxygen abundance and $R_{23}$ is double valued, requiring some assumption or rough a priori knowledge of a galaxy's metallicity in order to locate it on the appropriate branch of the curve. In this work, the [N II]$\lambda$6583/H$\alpha$ line ratio will be used to break the degeneracy of the $R_{23}$ relation (Denicoló et al. 2002). The division between the upper and the lower branch of the $R_{23}$ relation occurs around $\log([\text{N II}]\lambda 6583/\text{H}\alpha) \simeq -1.26$ $(12 + \log(\text{O/H}) \simeq 8.2)$.

### 3.3. P method

The $R_{23}$ method was used widely but $R_{23}$ abundances were found to be systematically larger than the $T_e$ metallicity. Pilyugin (2000, 2001) found that its error had two parts: a random error and a systematic error. The origin of this systematic error is the dependence of the oxygen emission lines on not only the oxygen abundance, but also on the other physical conditions (hardness of the ionizing radiation and a geometric factor). Pilyugin (2000, 2001) introduced the P method, another strong line metalloid indicator to overcome these problems. We use the expression of Pilyugin (2001) to calculate the abundance of oxygen in high metallicity regions $(\log([\text{N II}]\lambda 6583/\text{H}\alpha) > -1.26$, or $12 + \log(\text{O/H}) > 8.2)$:

$$12 + \log(\text{O/H})_P = \frac{R_{23} + 54.2 + 59.45P + 7.31P^2}{6.07 + 6.71P + 0.37P^2 + 0.243R_{23}} \quad (2)$$

and in low metallicity regions $(\log([\text{N II}]\lambda 6583/\text{H}\alpha) < -1.60$, or $12 + \log(\text{O/H}) < 7.95)$:

$$12 + \log(\text{O/H}) = 6.35 + 1.45\log R_3 - 3.19\log P, \quad (3)$$

where $R_3 = I([\text{O III}]\lambda\lambda 4959, 5007)/I(\text{H}\beta)$, and $P = R_3/R_{23}$.

### 3.4. N2 method

Both $R_{23}$ and $P$ metallicity are double valued. It is instructive to use one metallicity indicator to describe the whole metallicity with a single slope. The $N2 \equiv \log[I([\text{N{\sc ii}}]\lambda 6583)/I(\text{H}\alpha)]$ index was found to fulfill this requirement by Denicoló et al. (2002). A least squares fit to the data simultaneously minimizing the errors in both axes gives

$$12 + \log(\text{O/H}) = 9.12 + 0.73 \times N2. \quad (4)$$

The $N2$ indicator has advantages superior to the other metallicity indicators. The $N2$ vs. metallicity relation is monotonic, and the $N2$ line ratio does not depend on reddening corrections or flux calibration. These advantages make $N2$ indicators able to break the degeneracy of the $R_{23}$–(O/H) (in Sect. 3.2) and the $P$-(O/H) (in Sect. 3.3) relation.

### 3.5. O3N2 method

$O3N2 \equiv \log \{[I([\text{O{\sc iii}}]\lambda 5007)/I(\text{H}\beta)] / [I([\text{N{\sc ii}}]\lambda 6583)/I(\text{H}\alpha)]\}$ is another indicator that is monotonic. It was introduced by Alloin et al. (1979) and further studied by Pettini & Pagel (2004). Pettini & Pagel (2004) found that at $O3N2 \leq 1.9$, there appears to be a relatively tight, linear and steep relationship between $O3N2$ and $\log(\text{O/H})$. A least squares linear fit to the data in the range $-1 < O3N2 < 1.9$ yields the relation:

$$12 + \log(\text{O/H}) = 8.73 - 0.32 \times O3N2. \quad (5)$$

We use this expression to calculate the $O3N2$ oxygen abundance.

### 3.6. Bayesian method

It should be noted that the strong line methods, such as $R_{23}$, $N2$, $P$, $O3N2$, are empirical. The method based on physical model should be preferred rather than the empirical method. Besides the classic $T_e$ method, the Bayesian method is a good method based on a physical model. This method was proposed by Tremonti et al. (2004). The Bayesian method is based on simultaneous fits of all the most prominent emission lines ([O{\sc ii}], H$\alpha$, [O{\sc iii}], H$\beta$, [N{\sc ii}], [S{\sc ii}]) with a model designed for the interpretation of integrated galaxy spectra (Charlot & Longhetti 2001). We use Bayesian metallcities provided by the MPA/JHU group [7].

There are systematic differences between the Bayesian method and $T_e$ method (Shi et al. 2006, 2007). The origin of the difference between the Bayesian metallicities and $T_e$ metallicities have already been discussed by Yin et al. (2007). They found that for almost half of the sample galaxies (227 among 531 galaxies with $T_e$ measurements), Bayesian metallicities are overestimated by a factor of about 0.34 dex on average. They proposed that the overestimates of Bayesian metallicities may be related to the onset of secondary N enrichment in models. Another reason for the lower $T_e$ metallcities than Bayesian metallicities is that [O{\sc iii}]$\lambda$4363 emission line is biased by the very hot H{\sc ii} regions in each galaxy. Thus, the global average temperature might be overestimated by 1000-3000K, which results in the systematic underestimation of the oxygen abundance of 0.05-0.2 dex, as Nagao et al. (2006) proposed.

Generally speaking, all the methods should result in the same value of abundance for a given galaxy. This is not the case in practice. It is evident that the discrepancy is caused by problems both with models of HII regions and calibration. When comparing the numerical HII region models from Charlot & Longhetti (2001), which are at the base of the Bayesian abundances, with numerical models of other authors (Stasińska & Leitherer (1996), McGaugh's model (McGaugh 1991), CLOUDY (Ferland et al. 1998), or Kewley & Dopita (2002)), one will find that there is significant disagreement between those models. It is because the stellar evolutionary synthesis code and photoionization code used in these models are in continuous progress and improved. It is also a result of using different atomic data or different assumptions in these models. The significant disagreement between those models prohibits the present-day models from providing uniform oxygen abundances.

To help resolve this issue, one has to evaluate which is the most reliable oxygen abundance determination method to use when studying the metallicity of a galaxy. This will be the focus of the next section.

## 4. Occam's razor meets oxygen abundance indicators

### 4.1. Probability of a theory

Let us consider a set of N methods $\{H_i\}$, only one of which can be true, which we want to figure out. Then the probability of the k-th method, given the data D, is computed from Bayes' Theorem:

$$P(H_k|D) = \frac{P(H_k)\Delta(D|H_k)}{\sum_i p(H_k)\Delta(D|H_i)}$$

The prior probabilities $P(H_k)$ represent the investigator's degree of knowledge from previous measurements for the k-th method before seeing the data D. $\Delta(D|H_k)$ is a measure of how well $H_k$ fits the data. Bayes' Theorem states that the posterior probability for a certain method to be true is proportional to its prior probability assigned before seeing the data D and the degree that $H_k$ matches the data.

Then our assignment translates into measuring the prior probabilities $P(H_k)$ and $\Delta(D|H_k)$. Since prior probabilities are normalized over the parameter space,

$$\int P(\lambda_1, \lambda_2, ...)d\lambda_1 d\lambda_2 ... = 1,$$

the influence of prior probabilities on the posterior probability diminishes. $\Delta(D|H_k, \lambda_1, \lambda_2, ...)$ contributes significantly to the posterior probability. In practice, the prior hypotheses is quite unimportant, especially for good data (such as SDSS data). This is because good data have a good chance of supporting the correct method, even if the prior hypothesis is biased against another model. Only if the data are so bad or scant that they add little to our knowledge, will the posterior reflect the prior hypotheses.

For prior probabilities $P(H_k)$, we will assign equal prior probabilities to all methods, since we assume that we have no preference to any one of these methods in advance. As a result, to eliminate the influence of prior probabilities completely, we measure the mean likelihood of one method relative to another $\Delta_k$:

$$\Delta_k = \frac{1}{(2\pi\sigma^2)^{\frac{n}{2}}} exp(-\frac{1}{2\sigma^2} \sum_i lg(\frac{Z_0(D_i)}{Z(D_i)})^2).$$

Here our data is a sample of Gaussian distributions with mean $\mu$ and variance $\sigma^2$. $Z_0(D_i)$ is the standard metallcity for one

---
[7] See http://www.mpa-garching.mpg.de/SDSS/.

**Table 1.** The mean likelihood of one method relative to another. The likelihood are in unit of $10^4$.

|  | $T_e$ | Bayesian | $O3N2$ | $R_{23}$ | $P$ | $N2$ |
|---|---|---|---|---|---|---|
| $\Delta_k{}^1$ | - | -10.24 | 1.24 | -7.40 | 0.34 | -2.33 |
| $\Delta_k{}^2$ | -2.14 | - | -2.41 | 0.15 | -3.36 | -0.51 |
| $\Delta_k{}^3$ | 0.47 | -13.03 | - | -9.48 | 1.26 | -3.31 |
| $\Delta_k{}^4$ | -8.42 | 0.92 | -9.43 | - | -10.10 | -0.95 |

1: Ratio of the likelihood regarding $T_e$ abundance as observed abundance;
2: Ratio of the likelihood regarding *Bayesian* abundance as observed abundance;
3: Ratio of the likelihood regarding $O3N2$ abundance as observed abundance;
4: Ratio of the likelihood regarding $R_{23}$ abundance as observed abundance.

galaxy which is a 'temporary' supposition. $Z(D_i)$ is the metallicity derived by the ith-method. Table .1 gives the result of the mean likelihood of one method relative to another $\Delta_k$.

The results clearly show that the likelihood of the $T_e$, $P$ and $O3N2$ methods always have the same signs. This means that if any one method of the $T_e$, $P$ or $O3N2$ methods are (un)reliable, the other two methods will also be (un)reliable. It is a matter of course because the $P$ and $O3N2$ methods are calibrated by $T_e$-method. The likelihood of the Bayesian and $R_{23}$ methods always have same signs because both are calibrated by galaxy model. The likelihoods of the Bayesian method always have different signs than the classic $T_e$ method because there is a systematic difference between the Bayesian method and $T_e$ method (see Sect.3.6). In either case, likelihood of the $N2$ method is always negative using this dataset. This result may imply that the $N2$ method is an *unreliable* oxygen abundance determination method. An fundamental cause may lie in that the $N2$ method does not use oxygen emission line but only nitrogen emission line to calculate the oxygen abundance, which is obviously unsuitable.

### 4.2. The Bayesian Information Criterion($\mathcal{BIC}$)

Calculating probability of a theory in Sect. 4.1 does not give the reliability for a single method, so it is necessary to use another standard approach to test oxygen abundance determination methods, which is the Bayesian Information Criterion ($\mathcal{BIC}$).

Suppose we have two competing methods: $f(X_1; \theta_1 \ldots \theta_{m_1})$, $f(X_2; \phi_1 \ldots \phi_{m_2})$. We have a random sample: $X_1, X_2, \ldots X_n$. The likelihood functions for the two competing methods are: $L_1(\theta_1, \theta_2 \ldots \theta_{m_1})$ and $L_2(\phi_1, \phi_2 \ldots \phi_{m_2})$. The Bayesian Information Criterion ($\mathcal{BIC}$), which tests a hypothesis that one method fits the data better than the other, is defined as:

$$BIC = 2ln(\frac{L_1(\theta_1, \ldots, \theta_n)}{L_2(\phi_1, \ldots, \phi_n)}) - (m_1 - m_2) \times n. \quad (6)$$

Then we shall calculate the MLE's (Maximum Likelihood Estimates) $\widehat{\theta_i}$ and $\widehat{\phi_i}$ and then compute the estimated $\widehat{BIC}$:

$$\widehat{BIC} = 2ln(\frac{L_1(\widehat{\theta_1}, \ldots, \widehat{\theta_n})}{L_2(\widehat{\phi_1}, \ldots, \widehat{\phi_n})}) - (m_1 - m_2) \times n. \quad (7)$$

Using Eq.7, we can decide that the first method is superior to the second methods if it satisfies $\widehat{BIC} > 10$. According to hy-pothesis testing in this situation, a value of 10 is strong evidence that the model 1 is preferred over model 2.

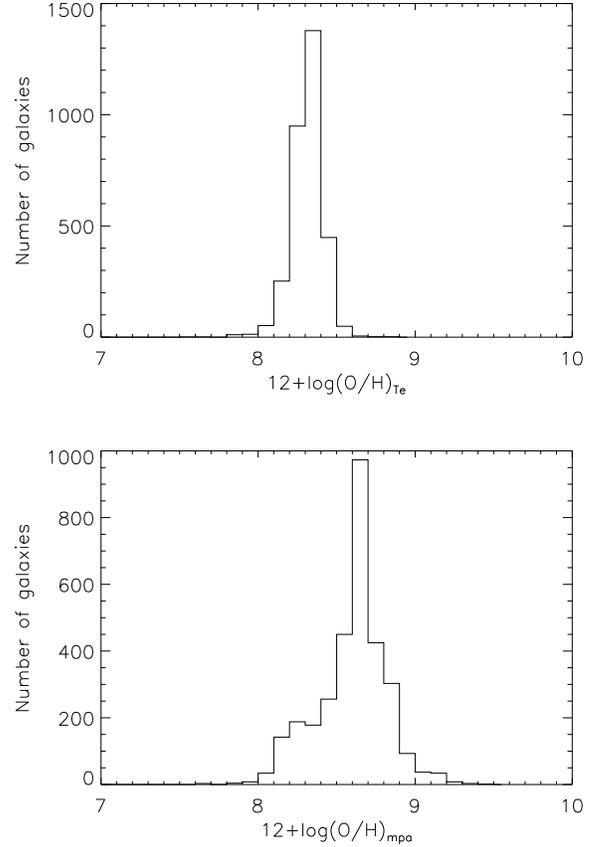

**Fig. 1.** Upper panel: The distribution for $T_e$ metallicity in our data sample. Lower panel: The distribution for Bayesian metallicity in our data sample.

Suppose our sample has a Gaussian distribution with mean $\mu$ and variance $\sigma^2$. This is a good approximation to the real distribution which is shown in Fig. 1. Note that the second term in Eq.7 always vanishes because the number of the parameters in our sample is 2 for all methods ($\mu$ and $\sigma^2$). As a result, the degree of one method fitting the data for the k-th method

$$L_k = \frac{1}{(2\pi\sigma^2)^{\frac{n}{2}}} exp(-\frac{1}{2\sigma^2} \sum_i (X_i - \mu)^2). \quad (8)$$

represents the validity of one method.

MLE for $\mu$ is $\overline{X}$. MLE for $\sigma^2$ is $S^2$. Then Eq. 8 is changed to:

$$L_k = \frac{1}{(2\pi\sigma^2)^{\frac{n}{2}}} exp(-\frac{n-1}{2}). \quad (9)$$

Table. 2 gives the result of the log likelihood for the validity of the $T_e$ methods and Bayesian method. To show the behavior of the $\widehat{BIC}$ for different types of galaxies, we calculate the likelihood for each subsample in bin of concentration index (C) and absolute luminosity in $R$ band ($M_r$).

Under the hypothesis testing framework, the title of "favorable method" is not normally bestowed unless $\widehat{BIC} > 10$. Table. 2 shows the consistent result for each type of galaxy. The log likelihoods for the validity of the $T_e$ methods are much higher

**Table 2.** The log likelihood for the validity of the $T_e$ methods and Bayesian method.

|  |  | $T_e$ | Bayesian |
|---|---|---|---|
| $\ln(L)^1$ | $C < 2.3$ | 903.6 | 166.5 |
| $\ln(L)^1$ | $2.3 < C < 2.6$ | 1341.1 | 333.8 |
| $\ln(L)^1$ | $C > 2.6$ | 808.8 | 212.2 |
| $\ln(L)^2$ | $-26 < M_r < -20.3$ | 121.0 | 49.0 |
| $\ln(L)^2$ | $-20.3 < M_r < -19.7$ | 151.6 | 74.6 |
| $\ln(L)^2$ | $-19.7 < M_r < -14$ | 282.9 | 92.3 |

1: We use $\widehat{BIC}$ method to find most reliable metallicity indicator in bin of C ;
2: We use $\widehat{BIC}$ method to find most reliable metallicity indicator in bin of $M_r$;

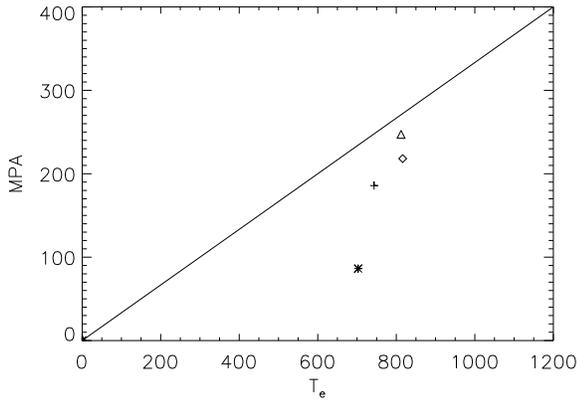

**Fig. 2.** The log likelihood for the $T_e$ method and the Bayesian method in bins of different ratio of signal to noise. Star denotes the log likelihood of the subsample of $S/N < 8$; Cross denotes the log likelihood of $8 < S/N < 10$; Diamond denotes the log likelihood of $10 < S/N < 13$; Triangle denotes the log likelihood of $S/N > 13$;

than the Bayesian method which makes $\widehat{BIC} \gg 10$. It is very strong evidence that $T_e$ method is more reliable than Bayesian method.

Though the decision is consistent, one should not trust the result blindly. Considering carefully and deeply on this matter will puzzle us because the Bayesian method is based on a physical model (Charlot & Longhetti 2001) which takes into account photoionization calculations and the various stages of the evolution of stellar populations, but the $T_e$ method is based on the much simpler model than Charlot & Longhetti (2001). Why is the credibility of Bayesian method not better than the $T_e$ method? The result of the Bayesian method elaborated in Sect. 3.6 is not convincing enough to solve this problem. To resolve this problem which cries for a solution, we plot the relationship between the log likelihood for the $T_e$ method and the Bayesian method in Fig. 2 in bins of different ratios of signal to noise (S/N).

Fig. 2 showes clearly that the credibility of the Bayesian method increases more quickly than $T_e$ method with the increase of S/N. We can safely predict that the credibility of the Bayesian method will exceed the credibility of $T_e$ method when the quality of the spectrum improves much better than SDSS. We anticipate that the next generation of telescopes can fulfill this mission.

### 4.3. Three Information Scoring Functions: AIC, BIC and ICOMP

While the previous section relies on Bayesian analysis, this section implements model selection based on information criteria. Traditional methods of model selection use hypothesis testing to draw some inference using two choices in a hypothesis test. The computed test statistic is compared with a threshold value to decide whether the researcher should accept or reject the hypothesis test. The preceding version of BIC actually performs a hypothesis test about which model is the correct one according to the computed likelihoods. The resulting value of BIC is then compared with a threshold value (we used 10) to see whether we should accept Model 1 or Model 2. We used this method in the previous section. By contrast, methods based on information theory do not rely on any hypothesis testing, but rather compute information scores to compare models. These methods try to balance the log likelihood term (Lack of Fit) with a penalty term to control the complexity of the model. This philosophy is consistent with Occam's razor. The following are the modern versions of these information scoring functions. The most parsimonious balance between these competing terms will show the minimum information score computed by these functions.

Information-based statistical analysis offers new ways to overcome problems associated with traditional methods. First proposed by Akaike (1973) and more fully developed by Bozdogan (1990, 2004), information scoring tries to find a balance between the lack of fit in model's log likelihood and a penalty term that controls the complexity of the model. This methodology resolves ambiguities associated with hypothesis test-based methods because values associated with hypothesis tests can be near the threshold of acceptance or rejection. By contrast, under the information scoring methodology, the model that achieves the lowest information score is the best to describe the system under study.

In our analysis, because the data all show univariate Gaussian distributions, we scored AIC, BIC and ICOMP under this assumption. The expression for the log likelihood is given by

$$-2logL(\Theta, Normal) = n\ln(2\pi) + n\ln(\sigma^2) + n \quad (10)$$

where $\sigma^2$ is the MLE of the variance and $n$ is the number of data points in the sample.

AIC is defined as minus 2 times the log likelihood plus 2 times the number of free parameters:

$$\begin{aligned} AIC(Normal) &= -2logL(\Theta) + 2m \\ &= n\ln(2\pi) + n\ln(\sigma^2) + n + 2 \times 2 \end{aligned} \quad (11)$$

where $logL(\Theta)$ is the maximized log likelihood and $m$ is the number of free parameters in the model. Likewise, BIC is given by:

$$\begin{aligned} BIC(Normal) &= -2logL(\Theta) + log(n)m \\ &= n\ln(2\pi) + n\ln(\sigma^2) + n + 2log(n) \end{aligned} \quad (12)$$

ICOMP is defined as:

$$\begin{aligned} ICOMP(Normal) &= -2logL(\Theta) + 2C_1(\Sigma_F) \\ &= n\ln(2\pi) + n\ln(\sigma^2) + n + 2C_1(\Sigma_F) \end{aligned} \quad (13)$$

**Table 3.** *AIC*, *BIC*, *ICOMP* for the $T_e$ method and the Bayesian method. The likelihood are in units of $10^3$.

|  | $T_e$ | *Bayesian* |
|---|---|---|
| *AIC* | -8.511 | -4.171 |
| *BIC* | -8.499 | -4.159 |
| *ICOMP* | -8.512 | -4.173 |

Instead of directly penalizing the number of free parameters, ICOMP penalizes the complexity of the Inverse Fisher Information Matrix. ICOMP is considered to be the most modern and consistent information scoring function (Bozdogan 2004). The Fisher Information Matrix for a univariate Gaussian distribution is given by:

$$\Sigma_F = \begin{pmatrix} \frac{n}{\sigma^2} & 0 \\ 0 & \frac{n}{2\sigma^4} \end{pmatrix} \quad (14)$$

The complexity component of ICOMP is:

$$C_1(\Sigma_F^{-1}) = \frac{q}{2} \log \left[ \frac{tr(\Sigma_F^{-1})}{q} \right] - \frac{1}{2} \log \left[ \det(\Sigma_F^{-1}) \right] \quad (15)$$

Here, $q = rank(\Sigma_F)$.

Using these expressions, we can compute the respective information scores for the different oxygen abundances.

In table 3, we give the result of these three scoring functions for the $T_e$ method and the Bayesian method. It shows the consistent result with table 2 and table 1 that the $T_e$ method is more reliable than the Bayesian method using the current data sample. The three scoring functions (*AIC*, *BIC*, *ICOMP*) agree with each other and all confirm that, according to the sample of computed oxygen abundances, the $T_e$ method fits the data better than the Bayesian method.

## 5. Conclusions

We have presented a large sample of spectroscopic measurements of H II galaxies from SDSS DR4 covering a wide range of metallicites ($7.5 < 12 + \log(O/H) < 9.0$). We have determined oxygen abundances for the sample, using different oxygen abundance indicators. We have studied the credibility of the different oxygen abundance indicators and obtained the following results.

1. $T_e$- method is the more reliable oxygen abundance determination methods than Bayesian metallcity under the existing telescope ability. We predict Bayesian metallcity will gradually become the most reliable oxygen abundance determination method when the S/N of the spectrum is much higher than nowadays.
2. $T_e$-, *P*-, *O3N2*-method are consistent with each other because *P*-, *O3N2*-method are calibrated by $T_e$-method. Bayesian, $R_{23}$- method are consistent with each other because both are calibrated by a galaxy model.
3. In either case, the *N2* method is an *unreliable* method.

*Acknowledgements.* This work is supported by the Chinese National Science Foundation (No. 10521001, and No. 10433010). Liang,Y.C are thanked for helpful suggestions. We are grateful to the AGN group at the Center for Astrophysics, University of Science of Technology of China for processing the SDSS spectra for continuum decomposition and line fitting using the spectral analysis algorithm developed by the group. Funding for the Sloan Digital Sky Survey (SDSS) has been provided by the Alfred P. Sloan Foundation, the Participating Institutions, the National Aeronautics and Space Administration, the National Science Foundation, the U.S. Department of Energy, the Japanese Monbukagakusho, and the Max Planck Society. The SDSS Web site is http://www.sdss.org/. The SDSS is managed by the Astrophysical Research Consortium (ARC) for the Participating Institutions. The Participating Institutions are The University of Chicago, Fermilab, the Institute for Advanced Study, the Japan Participation Group, The Johns Hopkins University, the Korean Scientist Group, Los Alamos National Laboratory, the Max-Planck-Institute for Astronomy (MPIA), the Max-Planck-Institute for Astrophysics (MPA), New Mexico State University, University of Pittsburgh, University of Portsmouth, Princeton University, the United States Naval Observatory, and the University of Washington.

## References

Akaike, H., 1973. Information theory and an extension of the maximum likelihood principle. In: Petrov, B.N., Csaki, F. (Eds.), Second International Symposium on Information Theory. Academiai Kiado, Budapest, pp. 267-281.
Alloin, D., Collin-Souffrin, S., Joly, M., & Vigroux, L. 1979, A&A, 78, 200
Bozdogan, H. (1990). On the Information-Based Measure of Covariance Complexity and Its Application to the Evaluation of Multivariate Linear Models, Communications in Statistics, Theory and Methods, A19 No.1,1990, pp. 221-278.
Bozdogan, H. (2004). Statistical Data Mining and Knowledge Discovery. Chapman and Hall/CRC, Boca Raton, Florida, 15-56.
Charlot, S, Fall, S. M. 2000, ApJ, 539, 718C
Charlot, S. & Longhetti, M. 2001, MNRAS, 323, 887
Contini, T., Treyer, M. A., Sullivan, M., & Ellis, R. S. 2002, MNRAS, 330, 75
Denicoló, G., Terlevich, R., & Terlevich, E. 2002, MNRAS, 330, 69
Dong, X.-B., Zhou, H.-Y., Wang, T.-G., Wang, J.-X., Li, C., & Zhou, Y.-Y. 2005, ApJ, 620, 629
Ferland, G. J., Korista, K. T., Verner, D. A., Ferguson, J. W., Kingdon, J. B., & Verner, E. M. 1998, PASP, 110, 761
Izotov, Y. I., Stasińska, G., Meynet, G., Guseva, N. G., & Thuan, T. X. 2006, A&A, 448, 955
Kauffmann, G., et al. 2003, MNRAS, 346, 1055
Kobulnicky, H. A., Kennicutt, R. C., Jr., & Pizagno, J. L. 1999, ApJ, 514, 544
Kennicutt, R. C. 1998, ARA&A, 36, 189
Kewley, L. J., & Dopita, M. A. 2002, ApJS, 142, 35
Kewley, L. J., Jansen, R. A., & Geller, M. J. 2005, PASP, 117, 227
Li, C., et al. 2005, AJ, 129, 669.
Liang, Y. C., Yin, S. Y., Hammer, F., Deng, L. C., Flores, H., & Zhang, B. 2006, ApJ, 652, 257
Lu, H., Zhou, H., Wang, J., Wang, T., Dong, X., Zhuang, Z., & Li, C. 2006, AJ, 131, 790
McGaugh, S. S. 1991, ApJ, 380, 140
Magueijo, J., & Sorkin, R. D. 2007, MNRAS, 377, L39
Nagao, T., Maiolino, R., & Marconi, A. 2006, A&A, 459, 85
Pagel, B. E. J., Edmunds, M. G., Blackwell, D. E., et al. 1979, MNRAS, 189, 95
Pettini, M. & Pagel, B. E. J. 2004, MNRAS, 348, L59
Pilyugin, L. S. 2000, A&A, 362, 325
Pilyugin, L. S. 2001, A&A, 369, 594
Pilyugin, L. S., & Thuan, T. X. 2005, ApJ, 631, 231
de Robertis, M. M., Dufour, R. J., & Hunt, R. W. 1987, JRASC, 81, 195
Shaw, R. A. & Dufour, R. J. 1995, PASP, 107, 896
Shi, F., Kong, X., Li, C., & Cheng, F. Z. 2005, A&A, 437, 849
Shi, F., Kong, X., & Cheng, F. Z. 2006, A&A, 453, 487
Shi, F., Zhao, G., & Liang, Y. C. 2007, A&A, accepted
Stasinska, G., & Leitherer, C. 1996, ApJS, 107, 661
Stasińska, G. 2005, A&A, 434, 507
Stasińska, G. 2006, A&A, 454, L127
Tremonti et al. 2004, ApJ, 613, 898T
Yin, S. Y., Liang, Y. C., Hammer, F., Brinchmann, J., Zhang, B., Deng, L. C., & Flores, H. 2007, A&A, 462, 535
York, D.G., Hall, P.B. et al, 2001, AJ, 122, 549